\documentclass[]{aa}
\usepackage{epsfig}
\usepackage{psfig}
\usepackage{graphics}
\usepackage{latexsym}
\begin{document}
\title{Spatial analysis of solar type III events associated with
narrow band spikes at metric wavelengths}
\author{G. Paesold\inst{1,2} \and A.O. Benz\inst{1} \and 
K.-L. Klein\inst{3} \and N. Vilmer\inst{3}}
\offprints{G. Paesold}
\mail{gpaesold@astro.phys.ethz.ch}
\institute{Institute of Astronomy, ETH Zentrum, CH-8092 Zurich, 
Switzerland \and Paul Scherrer Institute, W\"urenlingen und Villigen, CH-5232 
Villigen PSI, Switzerland \and
Observatoire de Paris, Section de Meudon, DASOP \& CNRS UMR 8645, 92195 
Meudon, France}\date{}
\titlerunning{Spatial analysis of type III bursts and metric radio spikes}
\authorrunning{Paesold et al.}
\abstract{
The spatial association of narrow band metric radio spikes with 
type III bursts is analyzed. The analysis addresses the question of a
possible causal relation between the spike emission and the
acceleration of the energetic electrons causing the type III burst.
The spikes are identified by the Phoenix-2 spectrometer (ETH Zurich)
from survey solar observations in the frequency range from 220 MHz to
530 MHz. Simultaneous spatial information was provided by the Nan\c{c}ay
Radioheliograph (NRH) at several frequencies. Five events were 
selected showing spikes at one or two and type
III bursts at two or more Nan\c{c}ay frequencies. The 3--dimensional 
geometry of the single events has been reconstructed by applying 
different coronal density models. As a working hypothesis it is 
assumed that emission at the plasma frequency or its harmonic is 
the responsible radiation process for the spikes as well as for the 
type III bursts. It has been found that the spike source location is 
consistent with the backward extrapolation of the trajectory of the type III 
bursts, tracing a magnetic field line. In one of the analyzed events, 
type III bursts with two different trajectories originating from the 
same spike source could be identified. These findings support the 
hypothesis that narrow band metric spikes are closely related to the 
acceleration region.
}
\date{Received 12 February 2001/ Accepted 19 February 2001}
\maketitle
\section{Introduction}
Millisecond narrow band radio spikes are structures in
the radio spectrum of the Sun forming a distinct class of flare
emission. The term 'narrow band, millisecond spikes' refers to short
(few tens of ms) and narrow band (few percent of the center frequency)
peaks in the radio spectrogram. They can be observed in the range of
0.3 to 8 GHz and occur mainly during the impulsive phase of a solar flare. 
Since the spike emission is often associated with enhanced hard X-ray
emission (Benz \& Kane \cite{benzkane1986}; G\"udel et
al. \cite{guedeletal1991}; Aschwanden \& G\"udel \cite{aschwandenguedel1992})
it is suspected that spikes are closely related to the actual
process of energy release in solar flares.\\
The short duration and the narrow bandwidth suggest a small source
size and therefore a high brightness temperature (up to $10^{15}$
K). Only a coherent mechanism can account for the emission
but none of the proposed mechanisms is generally accepted. Proposed
emission mechanisms are the electron cyclotron maser (e.g. Holman,
Eichler \& Kundu~\cite{holmanetal1980}, Melrose \&
Dulk~\cite{melrosedulk1982}, Aschwanden~\cite{aschwanden1990},
Robinson~\cite{robinson1991}) and upper hybrid or z-mode instabilities
combined with wave-wave coupling (e.g. Zheleznyakov \&
Zaitsev~\cite{zheleznyakovzaitsev1975}, Vlahos et
al.~\cite{vlahosetal1983}, Tajima et al.~\cite{tajimaetal1990} and
G\"udel \& Wentzel~\cite{guedelwentzel1993}).\\
A subclass of spikes originally found at metric wavelengths correlates 
with type III bursts (Benz et al.~\cite{benzetal1982}). They occur in 
clusters usually at frequencies slightly higher than the start frequency 
of the type III burst and are located in a dynamic spectrogram at the 
intersection of the extrapolated type III and the spike 
frequency. Although they may be slightly shifted in time (in a positive or 
negative direction) they significantly correlate with the extrapolated 
type III burst (Benz et al.~\cite{benzetal1996}).  
Since the spike location in the spectrogram is close to the 
extrapolated type III burst it is 
a reasonable assumption that spike and type III radiations are emitted 
at the same characteristic frequency.\\
This type of radio emission has been called 'metric spikes' in 
the literature (e.g. G\"udel \& Zlobec \cite{guedelzlobec1991}) and it 
is still unclear whether the spikes in the metric and the decimetric 
range belong to the same class of events.\\
Previously published spatially resolved observations of metric spike 
events (Krucker et al.~\cite{kruckeretal1995}; 
Krucker et al.~\cite{kruckeretal1997}) found the spike sources at high 
altitudes and suggest a model of energy release taking place in or close 
to the spike sources. Escaping beams of electrons cause the type III 
emission. Thus a scenario is conceivable in which the spikes may be 
a direct signature of the accelerator.\\ 
Using two-dimensional, spatially resolved data from the Nan\c{c}ay
Radioheliograph (NRH), it is possible to reconstruct the spatial 
configuration of the event and the relative position of the spike 
source with respect to the type III trajectory. The main purpose of 
this work is to test whether the geometry of the events 
supports the picture mentioned above.
\section{Instruments}
\subsection{The Phoenix-2 spectrometer}
Since 1998 the Phoenix-2 spectrometer operated by the ETH Zurich has been 
continuously 
recording radio data from sunrise to sunset.
The intensity and polarization are digitally measured by the frequency-agile 
receiver in the range from 0.1-4.0 GHz at a time resolution of 500 $\mu$s for
single channel measurements. The receiver consists of 4000 channels at
1, 3 or 10 MHz bandwidth, from which a reduced number can be freely 
selected. A full description of the instrument can be found in 
Messmer et al. (\cite{messmeretal1999}).\\ 
The data used herein were recorded at a bandwidth of 1 MHz and a time 
resolution of 100 ms in the frequency range from 220 to 550 MHz, chosen 
in collaboration with the Nan\c{c}ay Radioheliograph.
\subsection{The Nan\c{c}ay Radioheliograph (NRH)}
In July 1996 the Meudon Observatory began
daily observations with the improved 2D imaging radioheliograph in
Nan\c{c}ay (France). The instrument is described by Kerdraon \& Delouis
(\cite{kerdraondelouis1996}). Five frequencies in the range from
150-450 MHz can be observed simultaneously at a maximum number of
200 images per second. The antennas are organized in two perpendicular
arrays and digitally correlated on 576 channels resulting in
measurements of Stokes I and V. The observing bandwidth is 700 kHz. \\
Data were taken with a frequency configuration of 164.0, 236.6, 327.0, 410.5,
432.0 MHz at a mean time resolution of 125 ms.\\
For older data the NRH provides one-dimensional scans of the corona at
five frequencies, with both its east-west and north-south branches (The
Radioheliograph Group~\cite{heliogroup1993}). The integration time was 1~s. 
The NRH then observed at 164, 236.6, 327, 408 and 435~MHz.
\section{Observation and method}
Data from observations of Phoenix-2 were used to
identify type III events associated with metric spike emission.
The choice of events was restricted by the observing frequencies of
Nan\c{c}ay. At least one frequency was required to locate the spike 
emission, and the type III emission was required to be visible at least in 
two frequencies. The following four events were found in the 
recent data:\\ 
\\
{\em 99/09/27}: In the time range from 09:53.8 to 09:57.2 UT 
a type III burst group occurred and was accompanied by a cluster of 
metric spikes in the frequency range
300 to 370 MHz.\\
\\
{\em 00/05/29}: Type III burst group in the range from 220 - 420 MHz and
metric spike emission from 390 to 430 MHz in the time range from 12:25.00 
to 12:29.80 UT.\\
\\
{\em 00/06/21}: Metric spikes in the
frequency range from 340 MHz to 450 MHz, accompanied by a type III
burst group from 220 MHz to 450 MHz. The event time is 10:12.4-10:12.5 UT.\\
\\
{\em 00/06/22}: Type III burst group in the range from 220 MHz to 510 MHz,
whereas the metric spikes are located between 400-420 MHz at the
beginning of the event and between 430-500 MHz later on. The whole
event lasts from 13:41.5 to 13:44.5 UT. \\

In addition, three separate events on 92/08/18, already
published in Krucker et al. (\cite{kruckeretal1997}) using only 
VLA observations at 333 MHz have been investigated using data from the 
previous heliograph in Nan\c{c}ay. The events show metric
spikes in the frequency range from 310 MHz to 360 MHz and type III activity 
from around 300 MHz down to below 40 MHz. The time ranges for the events 
are 13:43:50 to 13:44:10 UT, 14:01:47 to 14:02:00 UT and 14:14:35 
to 14:14:40 UT, respectively.
\subsection{Phoenix-2 observations}
\begin{figure*}
  \includegraphics[width=16cm]{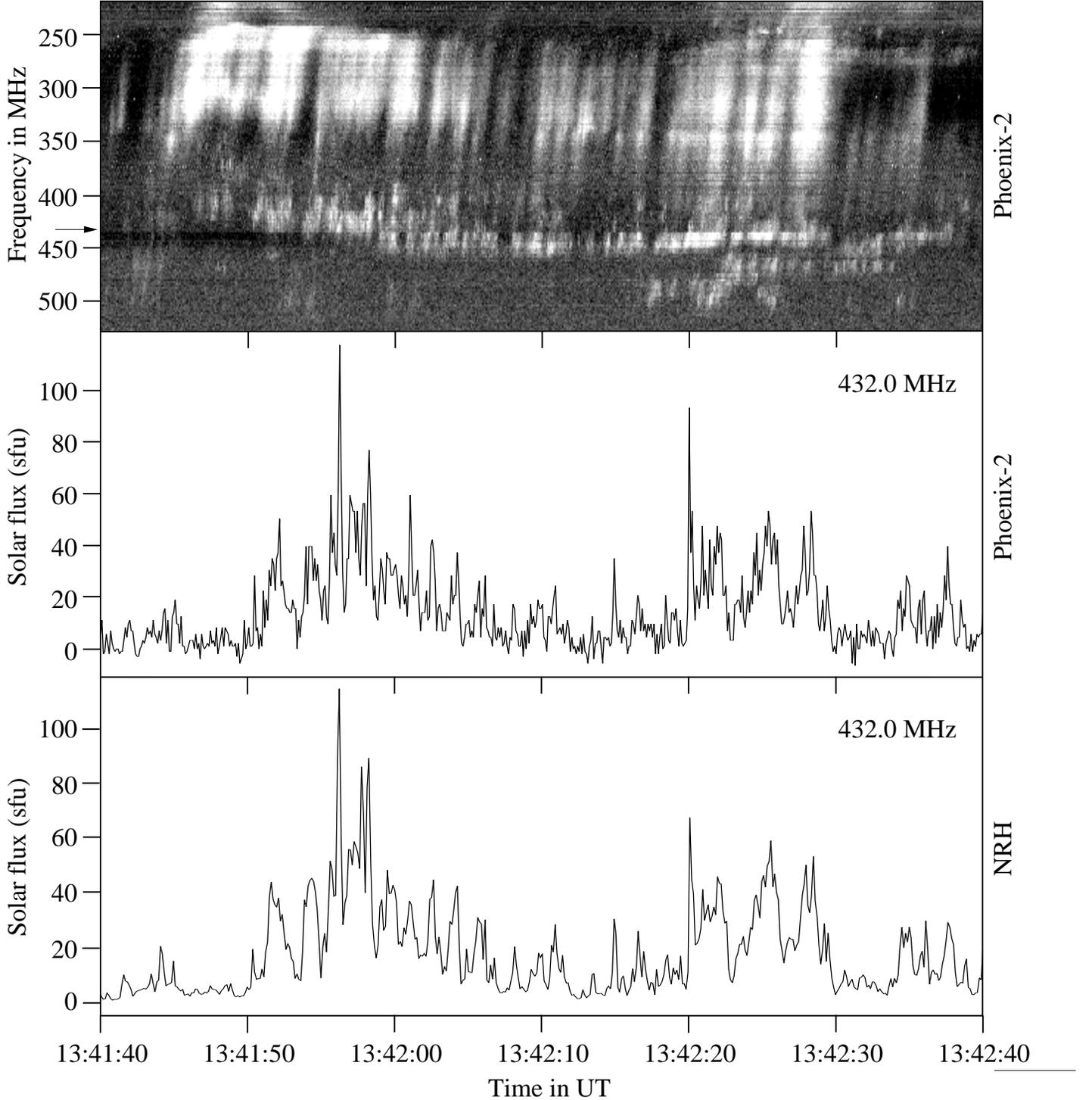}
  \caption[]{{\bf Top:} Spectrogram observed by Phoenix-2 on
    00/06/22. White regions correspond to enhanced flux; the 
    frequency axis
    is from top to bottom. {\bf Middle:} 
    Light curve of metric spikes recorded by the Phoenix-2 spectrometer
    at a single frequency (432.0 MHz). The arrow in the top panel
    indicates the position of 432.0 MHz. {\bf Bottom:} NRH light curve
    of metric spikes at the same frequency.}  
  \label{fig:1}
\end{figure*}
The dynamic spectrum from 220 MHz to 530 MHz of
the analyzed part of the 00/06/22 event is shown in Fig.~\ref{fig:1} (top). 
In the first half of the
the event (13:41:40 to about 13:42:08 UT) the metric
spike emission is mainly located in the range from 370 MHz to 470
MHz. At lower frequencies the drifting structures of type III emission
can be seen. In the second half, the spike emission is shifted to higher
frequencies (430 MHz to 510 MHz).\\
In the middle panel of Fig.~\ref{fig:1} a selected light curve at 432.0
MHz observed by the Phoenix-2 spectrometer is displayed. The data were
 recorded with a sampling period of 100 ms and an integration time
of 0.5 ms.  According to G\"udel \& Benz (\cite{guedelbenz1990}) the
expected mean time duration of a single spike event at 432.0 MHz is
$0.062\pm 0.004$ s. Therefore the single spike events are not
resolved in time. Following Messmer \& 
Benz (\cite{messmerbenz1999}), the minimum bandwidth of spikes is given
by $\sim 0.4\%$ of the center frequency yielding a value of 1.73 MHz
at 432.0 MHz. Although this value is close to the spectral resolution
of 1 MHz of Phoenix-2, the spikes are resolved in frequency and can be
well identified.\\
No polarization information is available from Phoenix-2
data for this event nor for the 00/06/21 and the 00/05/29 events 
since the spectrometer was run with a linear feed during this time.\\
The description of the observation mode of the old Phoenix
spectrometer for the 92/08/18 event can be found in Krucker et 
al. (\cite{kruckeretal1997}).  
\subsection{Nan\c{c}ay observations}
The bottom panel in Fig.~\ref{fig:1} shows the light curve at 432.0~MHz
of the same event observed by NRH. The data were recorded with an
integration time of 125 ms at a sampling rate of 125 ms. Since the
integration time of NRH is longer than of Phoenix-2, the noise
level of the dataset is lower.\\
The software used to analyze the 2D NRH data offers a source tracking
routine that allows the user to determine location and size of a
radio source in the 2D image of the Sun. The relevant source
was identified from Phoenix-2 observations by
simultaneously determining the flux in the NRH data
and comparing it to the full sun observation of Phoenix-2
(cf. Fig.~\ref{fig:1} middle and bottom). The size of the source is
determined by fitting an ellipse with minimal perimeter at half height
of the peak flux.\\
The error of the source centroid consists of two parts:
A discretization error and a statistical error. The first one stems 
from the NRH data being gridded.
The centroid position determined by the NRH software lies on a grid and 
the distance between two grid points is $\sim0.016 R_\odot$. Hence, 
there is a minimal error which is given by half the diagonal of a 
pixel $\sim0.011 R_\odot$.\\
The {\em statistical} uncertainty of the centroid position is given by the 
observed radius, $\Phi$, of the source times 
the noise to flux ratio. This ratio is constant according to the 
radiometer equation, and the resulting
error $\Delta$ can be written as
\begin{eqnarray}\label{eqn:1}
  \Delta = \frac{\sigma_\mathrm{max}}{F_\mathrm{max}}\Phi = 
           \frac{\sigma}{F_\mathrm{bg}}\Phi 
\end{eqnarray}
where the index $max$ refers to the peak values, $\sigma$ is the 
noise level at the background and $\mathrm{F}_\mathrm{bg}$ is the 
background flux. The noise can be determined from the background 
level at the source position before or after the event.
In most cases, the statistical error was smaller than the
discretization error, thus the latter one was used. Otherwise Eq.~\ref{eqn:1} 
yields the error bar.\\
In the case of the 92/08/18 event, the integration time in the NRH 
data is so long that the accuracy of the centroid determination is 
given by the pixel size. This has been verified in the following way:
1D scans were made at 164 and 327 MHz during five observing 
runs on Cyg A between 6 and 31 August 1992 and analyzed in the same way as 
the solar data. At 164 and 327 MHz, Cyg A appears as a simple 
source when observed with both the east-west and north-south array.
The measured centroid positions contain a random part and a slowly 
varying offset, which was evaluated by a polynomial fit.
The analysis yields a statistical uncertainty of $\sim1$ 
pixel (error bar) for both frequency channels and a systematic offset 
of one pixel in the interval of hour angles corresponding to the 
solar observations on 92/08/18.
\subsection{Association of spikes and type III bursts}
The identification of a single type III burst was mainly made by visual 
identification in the spectrogram. Only the stronger type III 
bursts of a group have been chosen for the analysis in order to ensure 
that they can be continuously traced back to the start frequency. In case 
of doubt the correlation function of 
two light curves at the relevant frequencies was computed and the burst 
was identified by correlation peaks. In case of more than one 
correlation peak, consistency with the drift rate between other 
frequencies was the criterion. However, it is still possible that two type 
III bursts with only a slight difference in drift rate crossover in 
the spectrogram and the associated sources do not 
belong to the same burst. As described in Sect.~\ref{results} the 
sources at each frequency of the analyzed events are very stable. 
Either each of the bursts in the 
group has a crossover with another type III on the ``wrong'' field line, 
which is not very probable, or the crossing type III 
burst occurs on the same field line and the source location is 
de facto not distinguishable from the source of the original burst.\\
By extrapolating the type III burst to the spike frequency, the associated 
spikes were identified. However, since the spikes can be slightly shifted 
in time with respect to the extrapolated type III burst (Benz et 
al.~\cite{benzetal1996}), a time window of about $\pm 0.2 $ s has to 
be used in order to take all possible associated spikes into account. 
The positions of all spikes inside the time window were included in 
the analysis.\\
Simultaneous emissions at two frequencies have been admitted if two of 
the observed frequencies lie in the spike range.
\section{Results}
\label{results}
\begin{figure}
  \resizebox{\hsize}{!}{\includegraphics{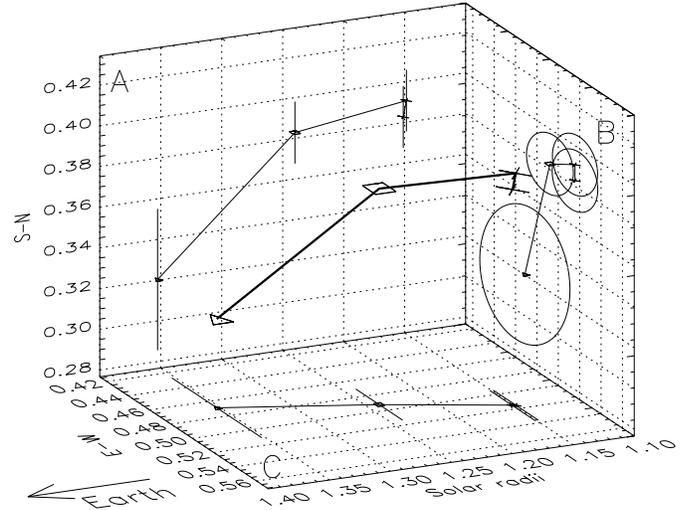}}
  \caption[]{3--dimensional view of one of the bursts in the 99/09/27
             event. The projection 
             on $A$ corresponds to the upper left panel in the plots of 
             Figs.~\ref{fig:3},~\ref{fig:4} and~\ref{fig:6}, the projection 
             on $B$ to the upper right and the projection on $C$ to the lower 
             right panel.}  
  \label{fig:2}
\end{figure}
The results for the four recent events (99/09/27, 00/05/29, 00/06/21, 
00/06/22) are presented in chronological order whereas the 92/08/18 events 
will be analyzed separately. In order to display three
dimensional situations in two dimensions, 
height projections on the solar equatorial plane and
the meridian plane which we define by the earth and the solar poles
have been chosen for representation in
Figs.~\ref{fig:3},~\ref{fig:4} and~\ref{fig:6} (see Fig.~\ref{fig:2}).
The upper right panel of each plot depicts the actual observations of 
Nan\c{c}ay as described in~\ref{subsec_obs} whereas the upper left and the 
lower right panel result from the 3--dimensional reconstruction 
described in Sect.~\ref{subsec_rec}. The symbols representing the observing 
frequencies are chosen as follows: $\star$~=~432.0~MHz, $+$~=~410.5~MHz, 
$\times$~=~327.0~MHz, $\Box$~=~236.6~MHz and $\triangle$~=~164.0~MHz.\\
The lines connecting the source centroids display how the sources are 
related to each other in the spectrogram. For reasons of visualization 
the observed source centroid positions in the 92/08/18 events have
been interpolated by 3--dimensional splines to smooth curves.\\
The sources do not vary significantly in the course of the event for all analyzed
events. Hence, for every event, one single representative 
burst out of the group has been plotted, although the analysis contains 
many more identified type III bursts.
\subsection{Observation}
\begin{figure*}
  \resizebox{\hsize}{!}{
    \vbox{\hbox{\psfig{file=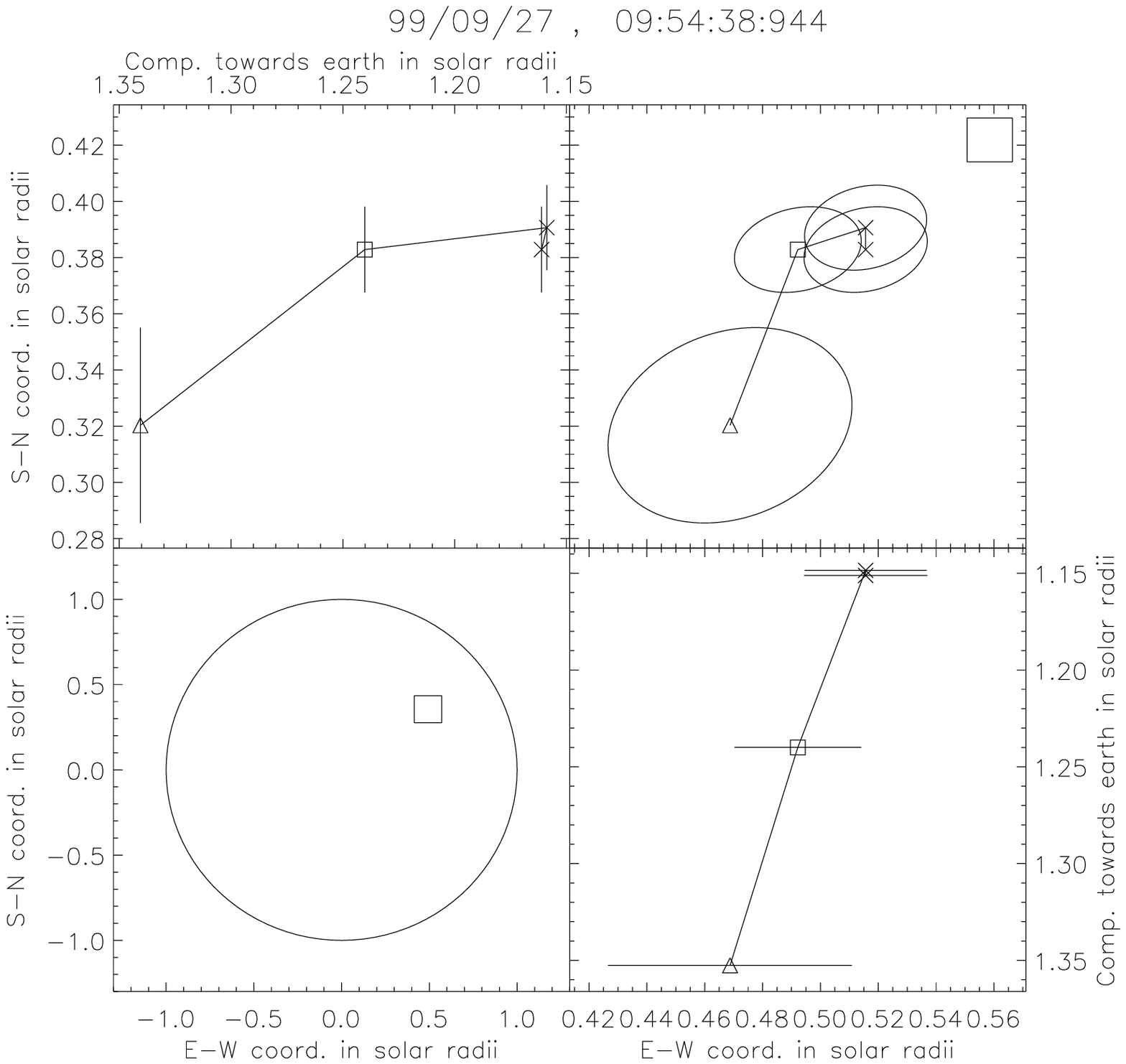}
                \psfig{file=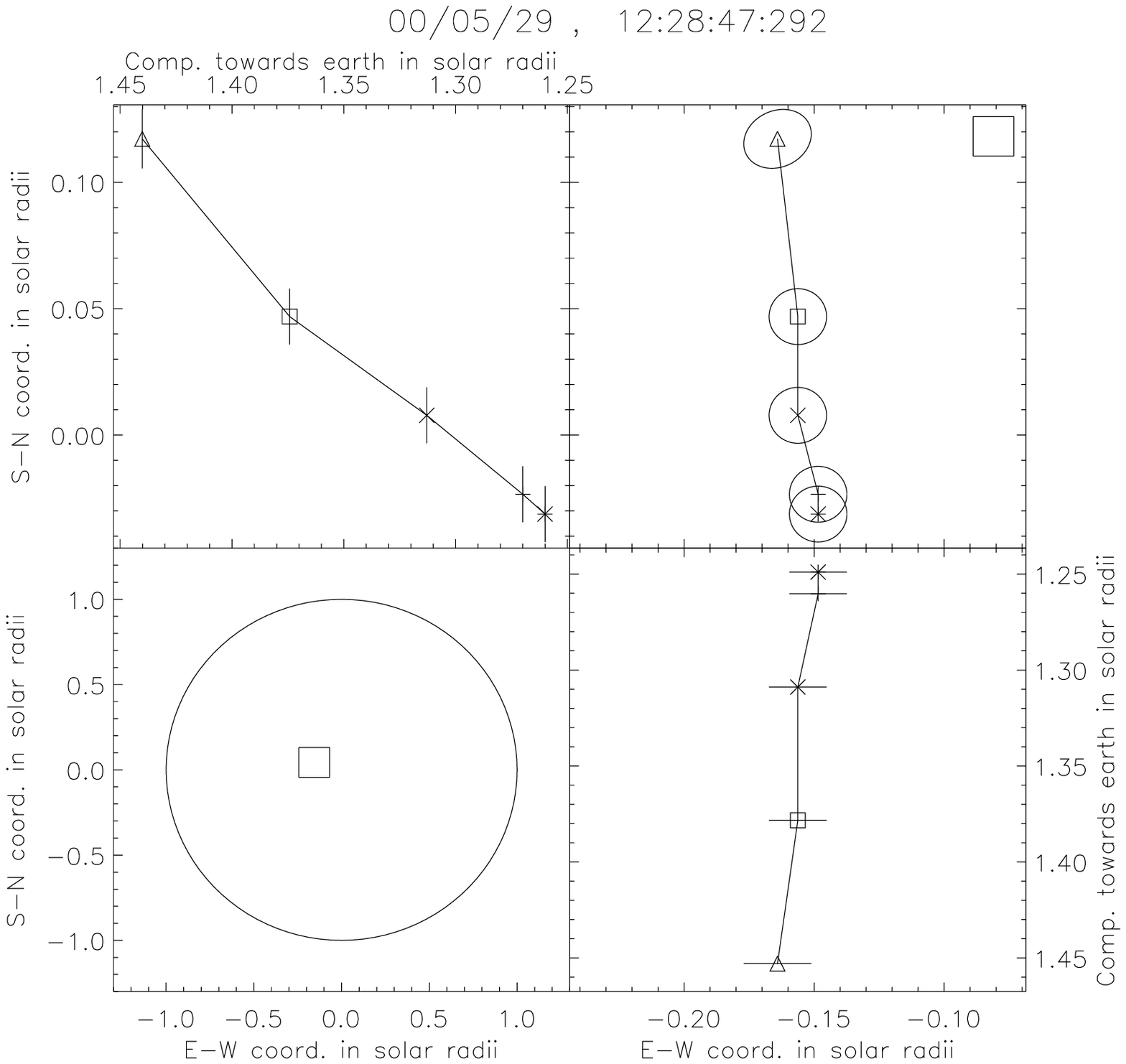}}\smallskip
          \hbox{\psfig{file=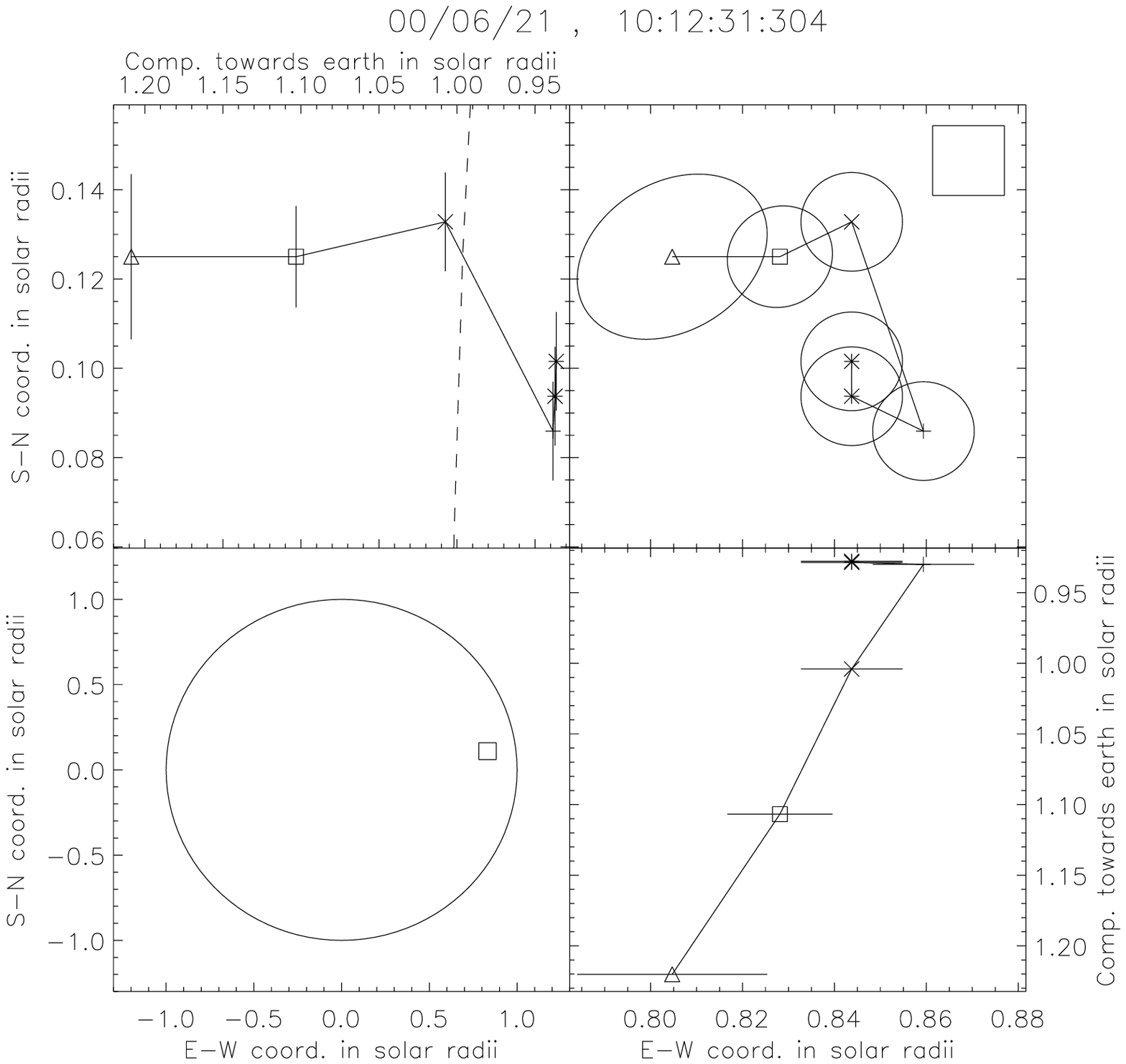}
                \psfig{file=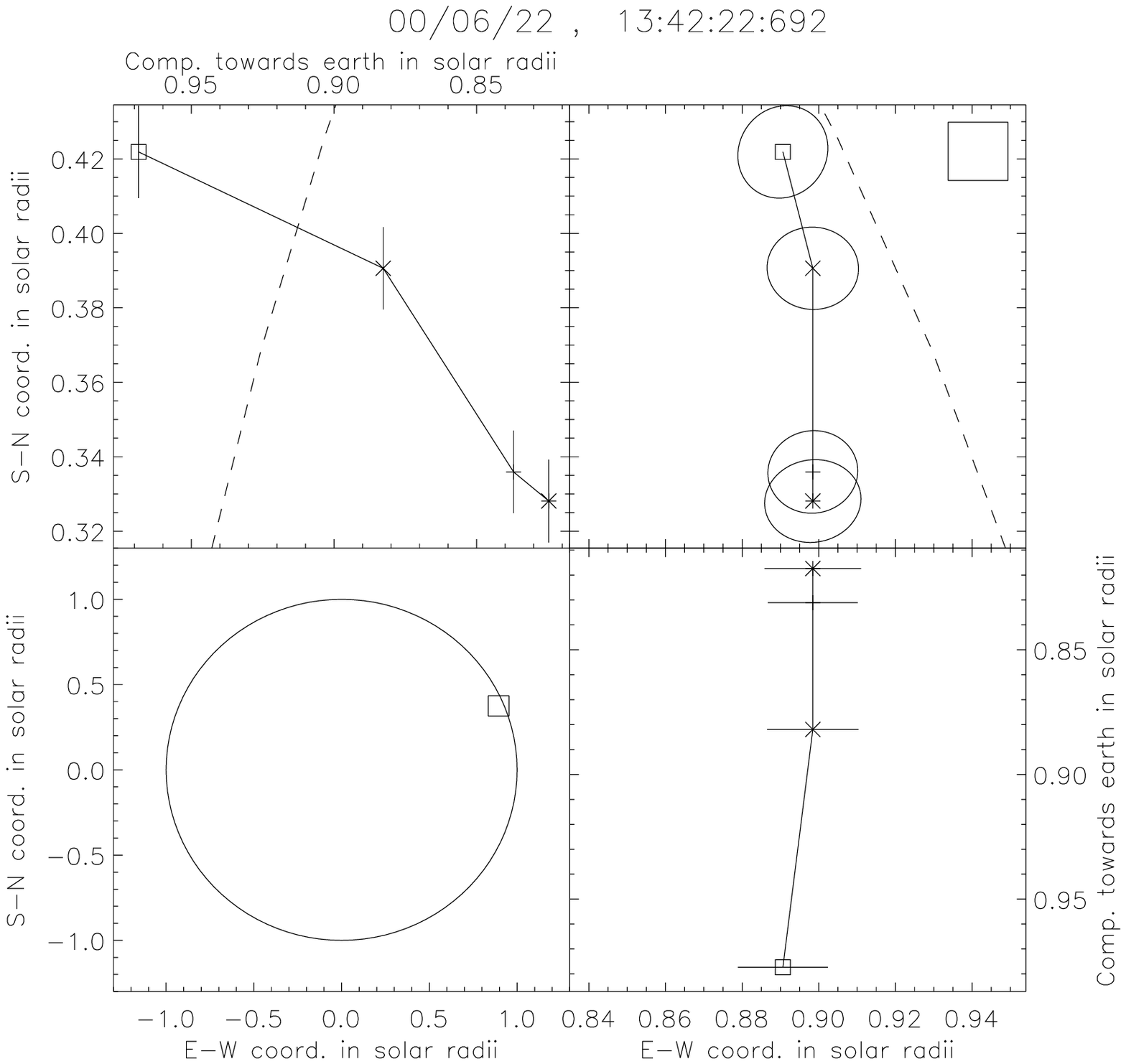}}}}
  \caption[]{A representative burst for each event is displayed. 
             {\sl Lower left:} global position of the event on the Sun. 
             The small square indicates the image size presented in the 
             {\sl upper right} quadrant, where the positions observed by 
             the NRH are given. The square in the upper right corner 
             represents the grid size of the NRH image. 
             The positional error of the source centroids are indicated
             by ellipses. {\sl Upper left:} Projection of the sources on
             the meridian plane (view seen from an observer 
             West of the 
             sources). {\sl Lower right} The sources in the projection
             on the equatorial plane, 
             showing the view of an observer North of the
             Sun (cf. Fig.~\ref{fig:2}).}  
  \label{fig:3}
\end{figure*}
\label{subsec_obs}
References to Figs.~\ref{fig:3},~\ref{fig:4} and~\ref{fig:6} in this 
section always refer to the upper right panel displaying the 
square shown in the overview (lower left panel).\\

{\em 99/09/27}:
Five type III bursts were identified at 164.0 MHz during this event. 
Two of the NRH frequencies (164.0 MHz, 236.6 MHz) lie in the range of 
the type III bursts and one (327.0 MHz) in the spike range. The event 
occurred on the first quadrant of the solar disc. The source 
positions at each frequency vary only within the error bars,
indicating a stable magnetic configuration during the event. 
Therefore it can be assumed that the bursts occurred practically on 
the same magnetic field lines. One representative burst of this 
group is depicted in Fig.~\ref{fig:3} (upper left plot).\\

{\em 00/05/29}:
Three type III bursts were identified at 164.0 MHz in this group 
occurring between 12:28:46 and 12:28:49 UT. All 
five observing frequencies of NRH could be used for this event. Three 
frequencies (164.0 MHz, 236.6 MHz, 327.0 MHz) lie in the range of 
the type III emission and two (410.5 MHz, 432.0 MHz) in the 
spike range. The spatial variation of the sources among the different
bursts is within the error bars and therefore the three bursts are 
assumed to have occurred approximately on the same magnetic field line. 
The upper right plot 
in Fig.~\ref{fig:3} displays one representative burst out of this group.\\

{\em 00/06/21}:
Between 10:12:28 and 10:12:44 UT six type III bursts were identified. 
Again all five frequencies of NRH could be used for the analysis. Three 
frequencies (164.0 MHz, 236.6 MHz, 327.0 MHz) lie in the range of 
the type III emission and two (410.5 MHz, 432.0 MHz) in the 
spike range. The event occurred close to the solar limb in the first quadrant. 
The sources at 164.0, 236.6, 327.0 and 410.5 MHz are stable within the 
error bars. The source at 432.0 MHz exhibits a larger spatial variation 
than observed in all other events. The type III source at 327.0~MHz is the 
most difficult one in the whole dataset to fit with a field line connecting 
the sources at all frequencies. The lower left plot in Fig.~\ref{fig:3} 
displays one representative burst.\\

{\em 00/06/22}: 
Three bursts out of this group have been analyzed. 236.6 MHz and 327.0 MHz
lie in the type III frequency range and 410.5 MHz and 432.0 MHz lie 
in the spike range. The spatial variation of the sources is negligible, and 
as in the other events the extrapolated type III trajectory intersects the 
observed spike source.
\subsubsection{92/08/18 event}

Three events of type III bursts with associated metric spikes were 
identified on 92/08/18 starting at 13:44 UT, 14:02 UT and 14:14 UT, 
respectively.\\

\begin{figure*}
  \resizebox{\hsize}{!}{
    \vbox{\hbox{\psfig{file=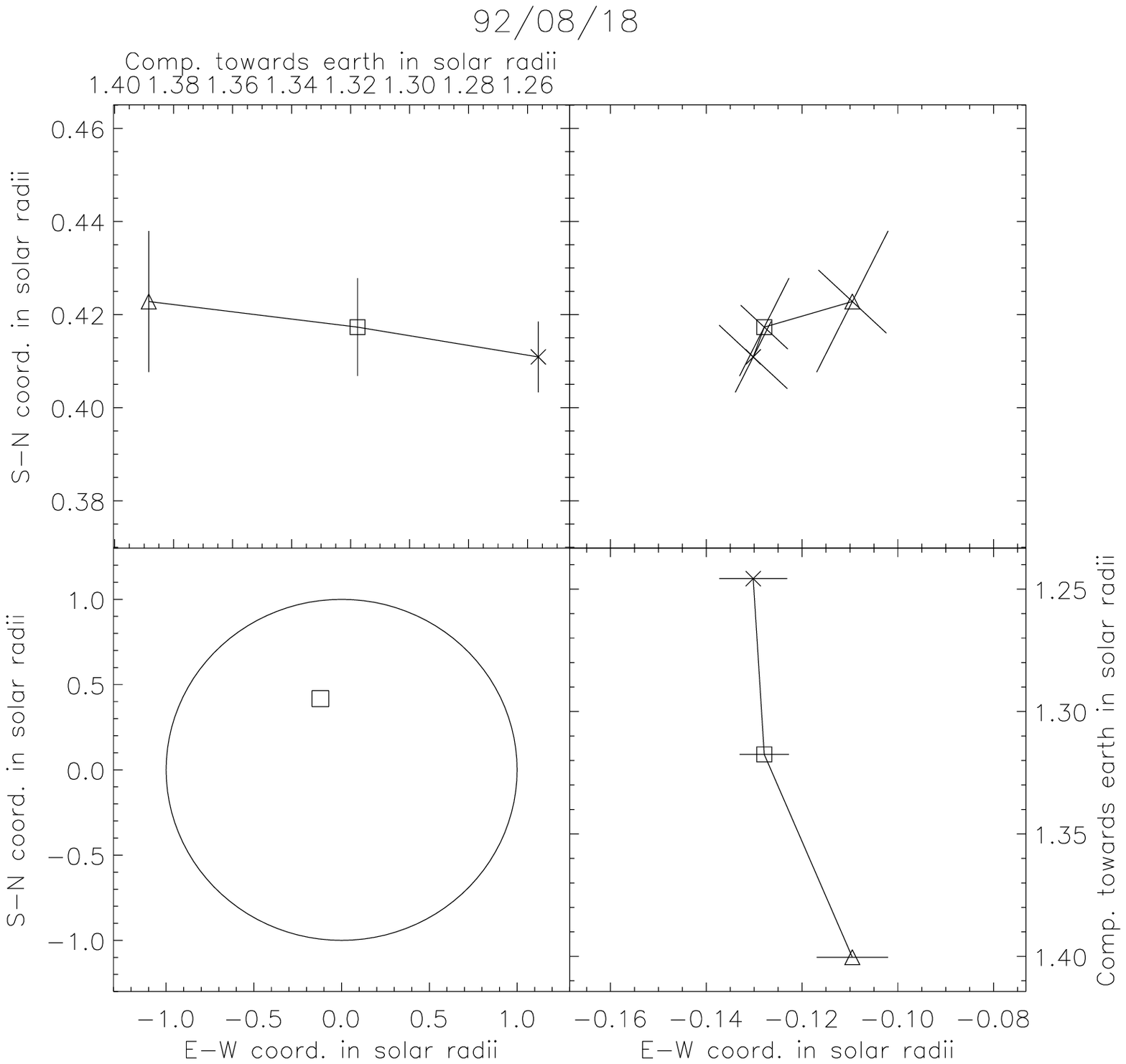}
                \psfig{file=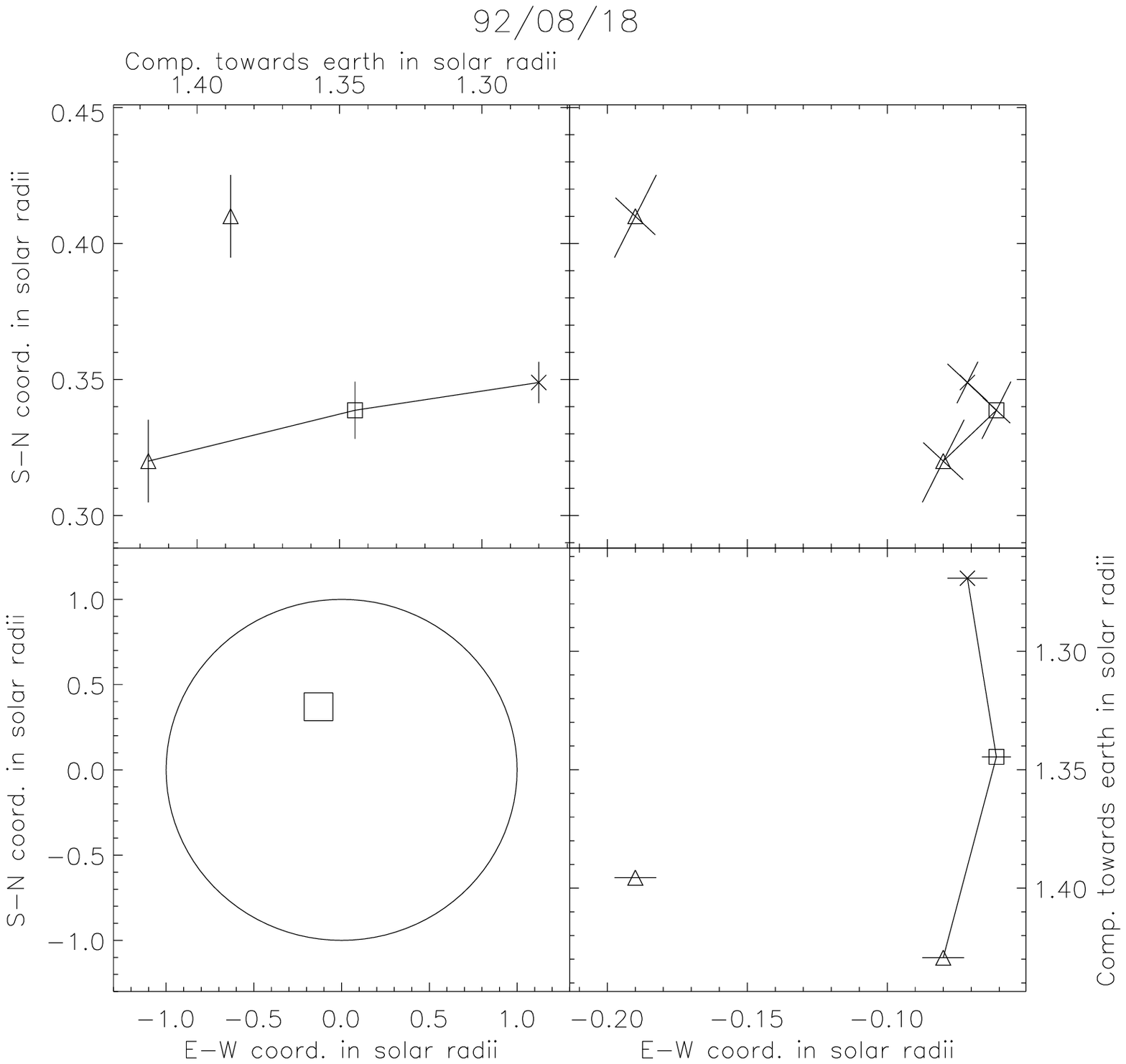}}}}
  \caption[]{{\bf Left:} A representative burst from the first type 
             III group at 13:44 UT, 92/08/18. {\bf Right:} Burst 
             in the second event at 14:02 UT, 92/08/18. The symbols 
             are the same as described in the caption of Fig.~\ref{fig:2}.}  
  \label{fig:4}
\end{figure*}
\begin{figure}
  \resizebox{\hsize}{!}{\includegraphics{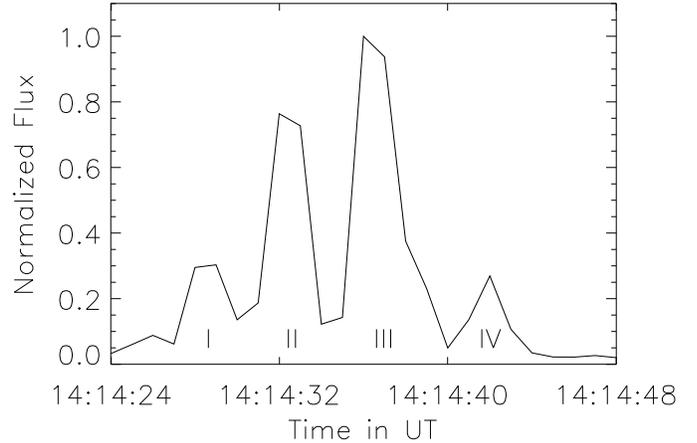}}
  \caption[]{Light curve of the 92/08/18 14:14 UT event at 164.0 MHz. The Roman
             numerals refer to the four identified type III bursts.}
  \label{fig:5}
\end{figure}
\begin{figure*}
  \resizebox{\hsize}{!}{
    \vbox{\hbox{\psfig{file=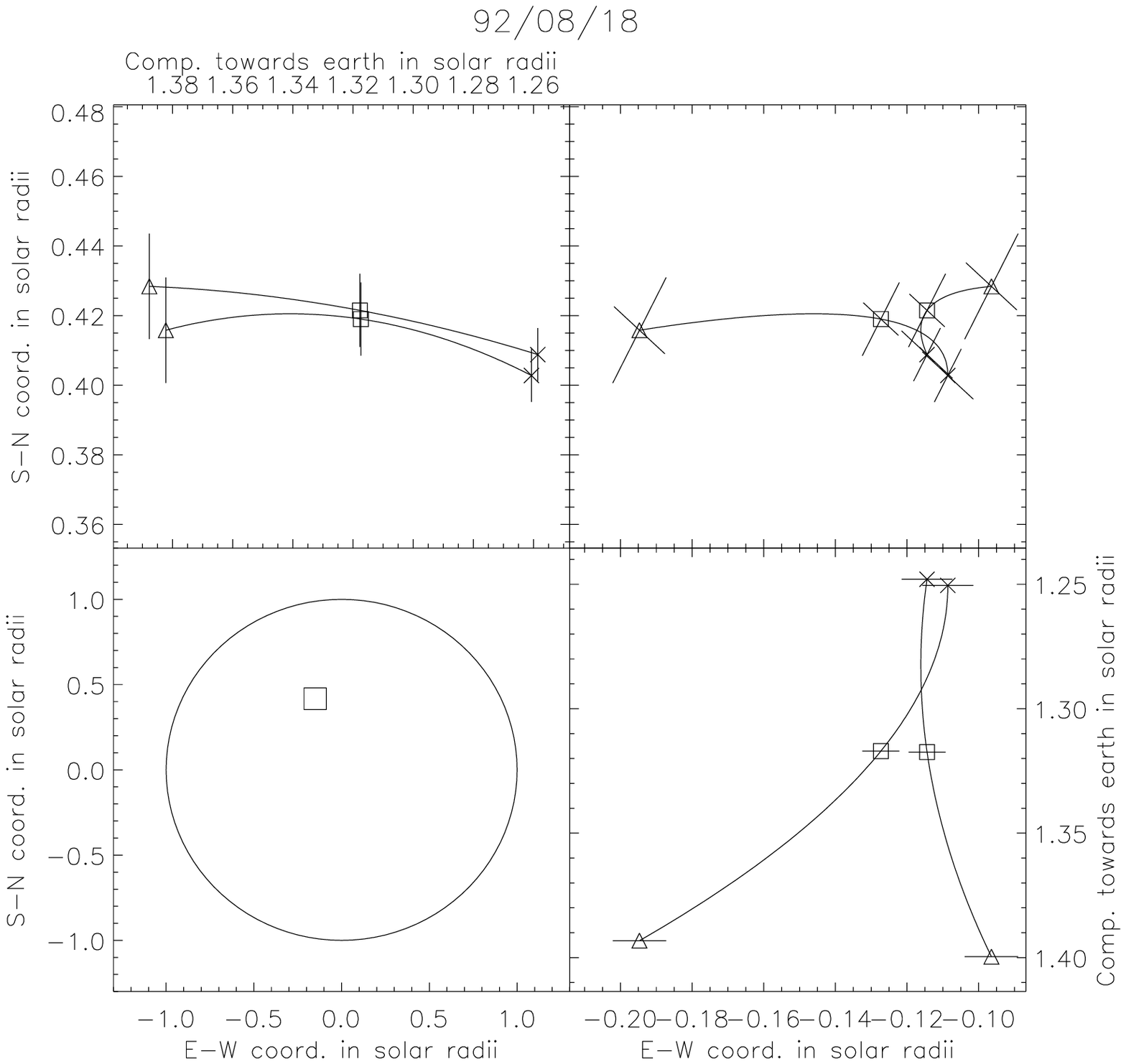}
                \psfig{file=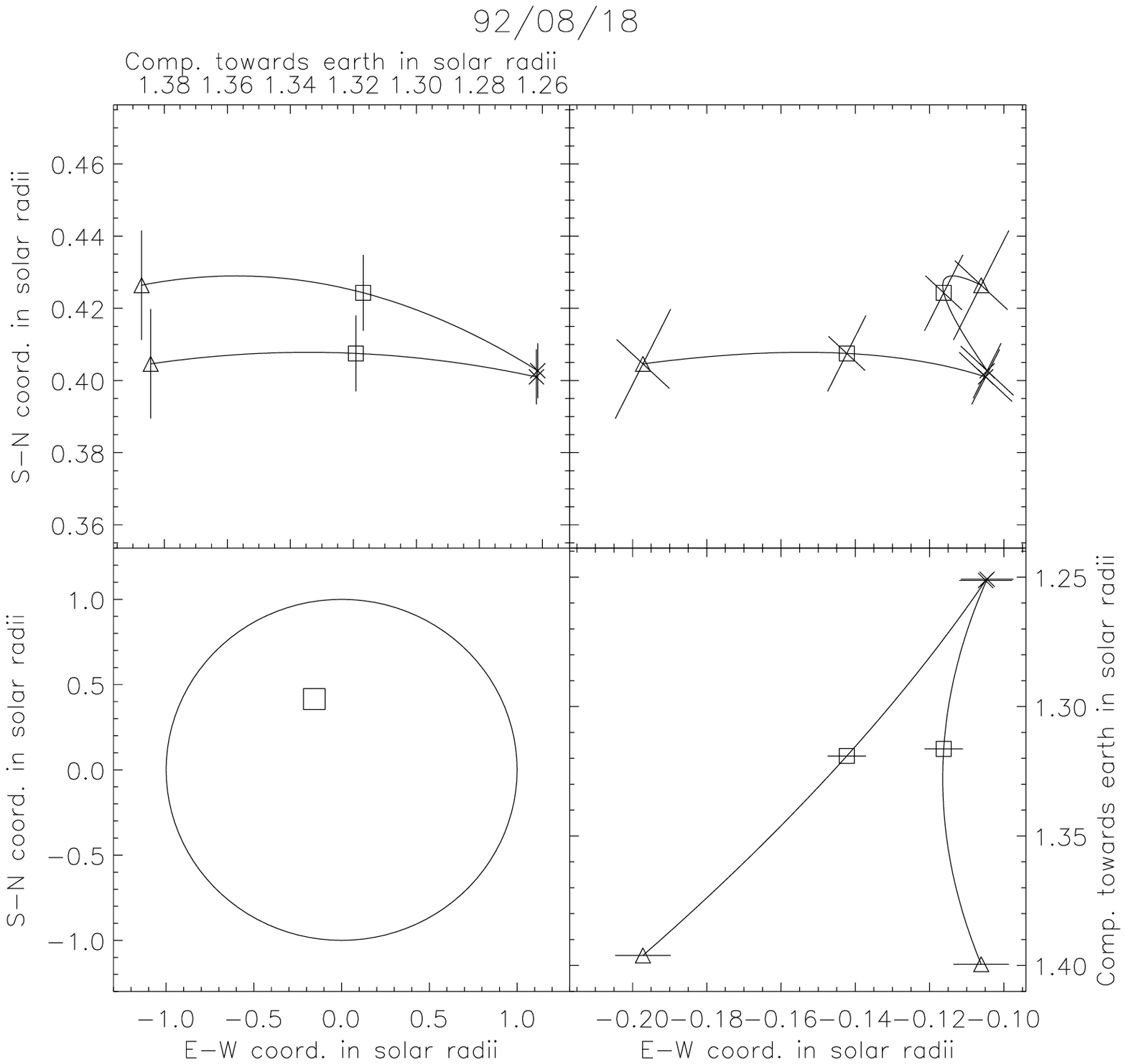}}}}
  \caption[]{{\bf Left:} Reconstructed trajectories of bursts I and II 
of the 92/08/18 14:14 UT event. The Roman numerals refer to the first 
two bursts identified in Fig.~\ref{fig:5}. {\bf Right:}
Burst III and IV of the 92/08/18, 14:14 UT event. In both plots 
the trajectories have been 3--dimensionally spline interpolated 
in order to make the distinction between error bars and trajectory easier. 
The symbols are the same as described in the caption of Fig.~\ref{fig:2}.}  
  \label{fig:6}
\end{figure*}
{\em 13:44 UT:} During the first event, three type III bursts were 
identified at 164.0 MHz. Two more NRH frequencies could be used for 
the analysis: 236.6 MHz in the type III range and 327.0 MHz in the spike 
range.\\ 
The scattering of the source centroid positions during the event lies 
within the error bars for each frequency. One representative burst is 
displayed in Fig.~\ref{fig:4}.\\

{\em 14:02 UT:} This event is of special interest since it exhibits
simultaneous sources at 164.0~MHz at different locations. Only one type 
III burst could be clearly identified in the spectrogram. 
At 164.0~MHz two sources were found: a stronger one more to the 
East and a weaker one more to the West. At 236.6~MHz only one source 
position could be clearly identified. This position is consistent 
with only one of the positions at 164.0~MHz. Due to the possible close 
positions 
of the sources and the time resolution of 1 s, a weaker second source at 
236.6~MHz would be difficult to detect.\\ 
At 327.0~MHz, in the spike range, one source was found. Hence, in 
Fig.~\ref{fig:4} on the right, both 164.0~MHz sources 
have been plotted whereas only one source at 236.6~MHz is displayed. 
The more probable trajectory in this configuration is depicted. 
Nevertheless, since the disconnected 164.0~MHz source is the 
stronger one and is without doubt part of the event, the situation 
is interpreted as two type III bursts propagating on different 
field lines, giving evidence of electrons being injected into 
different coronal structures from one single spike source.\\

{\em 14:14 UT:} The light curve at 164.0~MHz for 
this type III group is depicted in Fig.~\ref{fig:5}. Four bursts have been 
identified and labeled I-IV. Special attention has been given to this 
event for the following reason: Bursts I and IV occurred at a location 
significantly different from the position of bursts II and III. By 
analyzing the situation at 236.6~MHz, two sources were found with less, but 
still significant, spatial separation than at 164.0 MHz. At 327.0~MHz, a 
frequency lying in the spike range, the positions of the associated spike 
sources coincide.\\ 
Although the geometry is quite similar to the event at 14:02 UT, it exhibits a 
different situation since the bursts are consecutive and not simultaneous. 
Therefore, the identification and association of the sources was 
easier, and at every frequency the single bursts could be well 
identified.\\
\begin{figure}[h]
  \resizebox{\hsize}{!}{\includegraphics{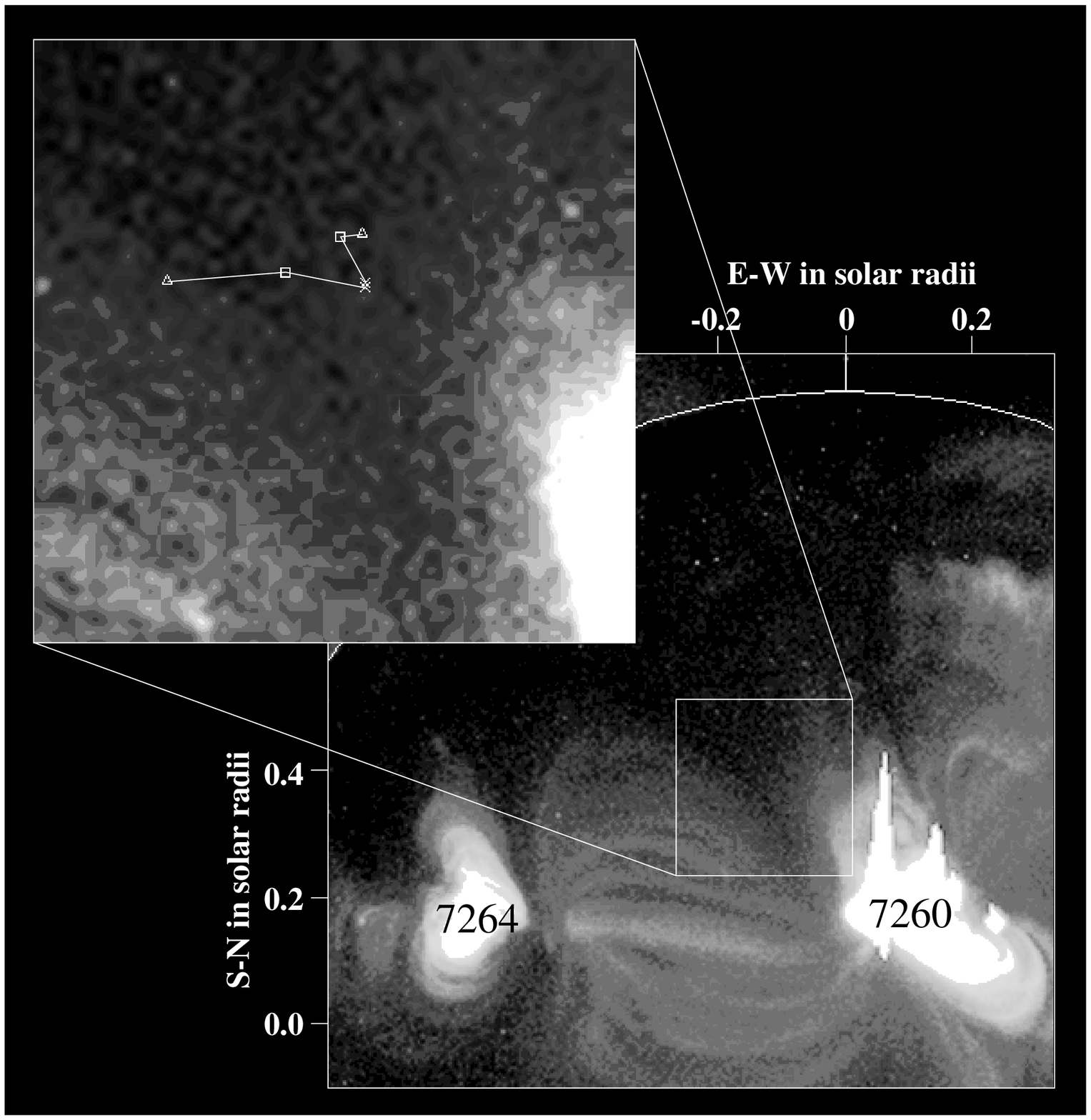}}
  \caption[]{Yohkoh SXT picture overlaid with radio sources. The SXT 
             image was taken with the Al.1 filter at 16:29:07 UT. The 
             integration time was 2.668 s. Two of the 92/08/18 event 
             radio bursts are displayed (see also Fig.~\ref{fig:6}, 
             right panel). The symbols are the same as described in 
             the caption of Fig.~\ref{fig:3}}
  \label{fig:7}
\end{figure}
The superposition of the radio centroid positions on the Yohkoh-SXT image 
(Fig.~\ref{fig:6}) suggests that the two electron beam trajectories 
inferred from the 
radio data correspond to magnetic field lines with different connectivities. 
The eastward oriented trajectory projects above loops that connect the 
trailing part of the active region (NOAA 7260) in the West with the 
leading part of an active region in the eastern hemisphere (NOAA 7264). 
The compact northward oriented trajectory is consistent with electron 
beams being guided by large-scale magnetic structures bending westward, 
possibly toward the leading part of AR 7260. 
The trailing part of AR 7260 had a complex magnetic polarity ($\delta$ spot) 
which produced several flares in August 1992 (cf. Leka et 
al.~\cite{lekaetal1996}) and is thus a plausible site for electron 
acceleration.
\subsection{Polarization}
With the exception of the limb event on 00/06/22, significant circular 
polarization was detected in all cases. The degree of polarization was 
moderate ($\sim$~10\%) for the type III emission, and much stronger (up 
to 90-100\%) for the spikes. The sense of the polarization was the same 
for spikes and type III bursts, consistent with the results found by 
Benz et al. (\cite{benzetal1996}) and references therein.
\subsection{Spatial reconstruction}
\label{subsec_rec}
\subsubsection{Coronal density models}
\begin{figure}
  \resizebox{\hsize}{!}{\includegraphics{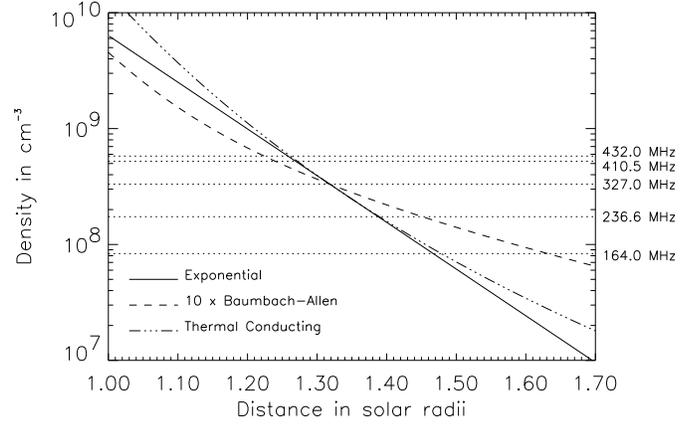}}
  \caption[]{Coronal density vs. distance from the Sun
    center. Three different models have been plotted.  The scale
    height and reference values of density and height for the exponential 
    and thermal conductivity model have been chosen
    to be $H_n=7.5\cdot10^9 cm$, $n_0=3.36\cdot10^8 cm^{-3}$ and
    $h_0=2.21\cdot 10^{10} cm$. The horizontal dotted lines indicate
    the densities corresponding to the observing frequencies of Nan\c{c}ay
    assuming emission at the harmonic of the plasma frequency.}
  \label{fig:8}
\end{figure}

Assuming emission either at the fundamental or harmonic of the plasma
frequency, the height of the radio source 
can be determined from a coronal density model $n_e(h)$ via
\begin{eqnarray}\label{eqn:2}
  \omega\approx\mathrm{a}\cdot\omega_p = 
  \left(\frac{4\pi\;e^2\;n_e(h)}{m_e}\right)^{1/2}\quad,
\end{eqnarray}
where $h$ is the height above the solar photosphere, and $a$ is about 
one or two for 
fundamental or harmonic emission, respectively.\\
Three different atmospheric models have been used and compared: 
An exponential atmosphere, the $10\times$Baumbach-Allen model
(Baumbach~\cite{baumbach1937}; Allen~\cite{allen1947}) and
an atmosphere in hydrostatic equilibrium with thermal 
conductivity (Lang~\cite{lang1980}). The latter two are given by
\begin{eqnarray}\label{eqn:3}
  n^\mathrm{BA}_e(h)&=&10\cdot1.55\cdot 10^8\;\left(\frac{h}
  {R_{\sun}}\right)^{-6}\times\nonumber\\
  &\times&\left[1+1.93\;\left(\frac{h}{R_{\sun}}\right)^{-10}\right]\;\;
  \mathrm{cm}^{-3}
\end{eqnarray}
for the Baumbach-Allen model and
\begin{eqnarray}\label{eqn:4}
  n^\mathrm{TC}_e(h)&=& n_0\left(\frac{h}{h_0}\right)^{-2/7}\times\nonumber\\
  &\times&\exp{\left[-\frac{7\;h_0}{5\; H_n}\left\{1-\left(\frac{h}{h_0}
  \right)^{-5/7}\right\}\right]}
\end{eqnarray}
for the thermally conducting corona. $n_0$ and $h_0$ are reference values 
for density and according height, $H_n$ is the scale height, 
corresponding to the scale height of the exponential density model. 
The reference values were chosen according to observations of 
Trottet et al.~(\cite{trottetetal1982}) and 
Suzuki \& Dulk~(\cite{suzukidulk1985}). Sources at 160 MHz have been 
found at heights of about 0.5 solar radii above the photoshpere 
which is reproduced by the model under the assumption of harmonic 
emission.
The resulting density from all three models is depicted in 
Fig.~\ref{fig:8}.\\
Although the solar corona is not expected to have a spherically symmetric 
density distribution, the models are considered to apply to single magnetic 
flux tubes which are supposed to be the type III guiding structures in 
the corona. Being anchored in the parent active regions, the flux tubes can
exhibit curvature and significantly deviate from radiality. Nevertheless, 
the density within an individual flux tube can be assumed to follow a model 
such as described above.\\ 
The 3--dimensional position of the radio sources is assumed to be at 
the intersection of the line of sight
and the sphere defined by the density model (Eq.~\ref{eqn:2}). As all 
the sources were observed on the disc, there
was no ambiguity. Propagation effects and their impact on the source 
positions 
are discussed in Sect.~\ref{discus}.\\
If not mentioned otherwise, the exponential density model and $a=2$ has 
been used in the following. It is still controversial whether structureless 
type III bursts (i.e. no fundamental--harmonic pairs) are emitted at the 
fundamental or the harmonic of the plasma frequency. This question shall 
not be discussed here and we refer the reader to the review of Suzuki \& 
Dulk~(\cite{suzukidulk1985}) and references therein.\\
The results of the 3--dimensional reconstruction are shown in the 
upper left (side-view) and lower right (top-view) panels of each plot 
in Figs.~\ref{fig:3},~\ref{fig:4} and~\ref{fig:6}. The component towards 
Earth is the projection of the radio source's height on the axis 
perpendicular to the plane of the sky (cf. Fig.~\ref{fig:2}).\\
It is the most important result that in all cases analyzed here, the observed 
locations of the spikes coincide in a smooth and natural way with 
the expected position of radio emission at the corresponding frequencies 
from extrapolating the type III trajectory to lower altitudes.
\section{Discussion}
\label{discus}
In the following we discuss the findings in the previous section 
addressing the
3D reconstruction of the bursts, including possible influence of 
radio wave propagation 
effects, and the interpretation of spikes being a signature of 
the accelerator. 
\subsection{Source locations}
It is obvious that the reconstructed source locations depend on the 
chosen coronal density model in terms of absolute heights. The same is 
the case for the choice of emission at the harmonic rather than at the 
fundamental of the plasma frequency. However, the relative positions 
which are of major interest in this work are not altered by changing 
either the atmospheric models nor the characteristic emission 
frequency. The trajectories may be stretched and shifted in height 
but the topology of the burst remains the same.\\
The difference in polarization degree between type III (moderate) and 
spikes (strong) may indicate another emission frequency for the spikes 
than occurs for the type III burst (e.g. fundamental for spikes and harmonic for 
type III). This 
would shift the spike source to lower altitudes with respect to the 
type III burst without changing the burst topology. The analysis of 
Benz et al.~(\cite{benzetal1996}) shows that the spikes correlate with 
the extrapolated type III burst to higher frequencies (i.e. the spike 
frequency). 
In case of significantly different emission frequencies for spikes 
and type III burst, a systematic time offset would be expected.\\
A common feature of all analyzed events (Figs.~\ref{fig:3},~\ref{fig:4} 
and~\ref{fig:6}) is a bending in East -- West direction of the 
trajectories towards the line of sight to the observer, 
independent of the location on the solar disc. In North -- South 
direction the data also 
exhibit non-radiality, but no general trend of deviation was observed. 
Besides this result being an indication 
for curved magnetic field structures, there are propagation effects that 
could produce an apparent bending of the magnetic field lines:
\begin{itemize}
\item[--] Refraction in the corona can shift the apparent source location.
\item[--] The radio emission may be scattered in the corona during 
propagation to the observer and therefore can produce an apparent 
image of the source at a location different from the true position 
(Arzner \& Magun~\cite{arznermagun1999} and references therein).
\item[--] Type III emission may be beamed along the trajectory of the 
electrons, selecting only those type III bursts propagating 
towards the observer within the beaming cone (Caroubalous et 
al.~\cite{caretal1974}; Caroubalous \& Steinberg~\cite{carsteinberg1974}). 
\end{itemize}
All events have been analyzed by overlaying EIT images, as shown in 
Fig.~\ref{fig:7} for the 92/08/18 event with an SXT image. 
Comparable indications of magnetic field structures have been found 
that can explain at least part of the shift of the projected type 
III positions towards the observer by the guiding magnetic field. 
E.g., on 00/05/29, the only other event near central meridian, the 
observed configuration is consistent with sources in an open flux 
tube that is anchored in the leading part of the active region as 
seen by EIT and is part of the field lines which project northward 
onto the disk. Projection effects make association with active regions 
difficult in the event of 00/06/20. Nevertheless, the radio sources 
on 00/06/21 have virtually the same north-south coordinate at all 
frequencies, and the more they are shifted to the east, the 
lower the frequency. This is expected for an east-westward extending 
flux tube anchored in the leading part of the underlying active region, 
and this interpretation is consistent with the EIT image. \\
We thus believe that the observations lend plausibility to the 
assumption that the radio source positions are mainly affected 
by the magnetic field structure and to a lesser extent by 
propagation effects. This conclusion is also supported by the work of 
Pick \& van den Oord (\cite{pickvandenoord1990}) and references therein.
\subsection{Spike location and acceleration}
\begin{figure}[h]
  \resizebox{\hsize}{!}{\includegraphics{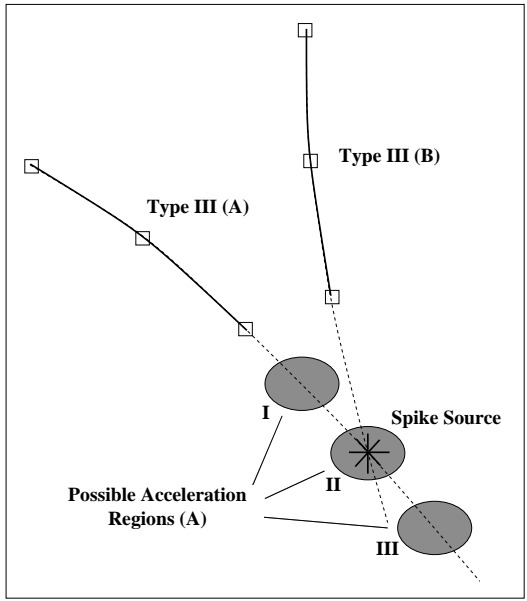}}
  \caption[]{Sketch of possible locations of the acceleration region 
             (I, II, III) with respect to a type III burst (labeled A) 
             and an associated spike source (asterisk). A second type 
             III (labeled B) is displayed in case of two simultaneous 
             bursts.}
  \label{fig:9}
\end{figure}
To inject electrons on the field line guiding the type III 
burst, the acceleration region must be located close to or on the 
field line itself. Fig.~\ref{fig:9} displays a sketch depicting 
possible locations of acceleration with respect to a type III burst (A) 
and a spike source. A priori, there are three different positions for 
the acceleration region consistent with the present observations: 
it lies between the type III burst and 
the spike source (case I), below the spike source (case III) or 
coincides with the spike source (case II). Assuming that the 
spike emission is caused by the same acceleration event, location I can be 
excluded by analyzing spectral radio observations. For the acceleration 
region to lie in between the type III and the spike source, the 
time of the actual acceleration must lie within the time interval defined by 
the intersection of the extrapolated type III with the spike frequency 
and the start of the type III emission.
A systematic time delay of the spike source with respect to the 
acceleration event must be observed, caused by the travel distance 
of the electrons generating the spike emission. 
According to Benz et al. (\cite{benzetal1996}) the 
spikes correlate with the intersection of the extrapolated 
type III and the spike frequency itself and no systematic delay was found. 
This analysis supports locations II and III as potential region of 
acceleration and excludes location I in Fig.~\ref{fig:9}.\\
The situation of two type III bursts (labeled A and B in Fig.~\ref{fig:9}) 
associated with a single spike source, 
as was found in two of the 92/08/18 events, suggests location II as a 
possible acceleration region. Position III is only consistent if the 
field lines meet in position II and continue in parallel to position III.
\section{Conclusions}
\label{conc}
In all analyzed events the spike sources are always located at 
positions coinciding with expected locations from extrapolated type III 
trajectories to lower altitudes. These observations thus strongly support 
a model for radio spikes occurring in the course of type III beam 
propagation or near its origin, consistent with independent spectrogram 
observations (Benz et al. 1996). They add further evidence for spikes 
being a signature of the mechanism accelerating electron beams that 
cause type III bursts. This appears to be the simplest interpretation 
(cf. Fig.~\ref{fig:9}).\\ 
This property is supported by the results of the 92/08/18 observations, 
where in two events simultaneous or consecutive type III bursts on different 
magnetic field lines originated in the same spike source. Energetic electrons 
appear to be injected into different and diverging coronal structures 
from one single position. Such a diverging magnetic field geometry is 
the standard ingredient of reconnection. These observations are consistent 
with the hypothesis that metric spikes may be a signature of particle
acceleration.\\
Earlier imaging investigations on metric spikes associated with type III 
bursts and their interpretation (Krucker et al.~\cite{kruckeretal1995}; 
Krucker et al.~\cite{kruckeretal1997}) can now be compared to these 
additional observations. They proposed a scenario of energy release at 
high altitude with up- and downward moving 
energized electrons. The upward moving electrons produce type III bursts 
while propagating along open field lines and the downward moving part loses 
its energy to the lower corona, transition region or upper chromosphere. 
The radio emission of electron beams moving downward from coronal 
acceleration sites has occasionally been detected (e.g. Klein et 
al.~\cite{kleinetal1997}), 
but no signature was seen in the observations presented here. 
High-sensitivity observations are necessary to investigate the
processes below the spike source, i.e. above the spike frequency band. 
\begin{acknowledgements}
We thank Christian Monstein, Michael Arnold and Peter Messmer (ETH Zurich) 
for helping to run the Phoenix-2 observations. The evaluation of the 
Phoenix spectrograms was made using software developed by A. Csillaghy. 
The work at ETH Zurich is financially supported by the Swiss National 
Science Foundation (grant No. 2000-061559.00).\\
The Nan\c{c}ay Radio Observatory is funded by the French Ministry of
Education, the CNRS and the R\'egion Centre.
\end{acknowledgements}

\end{document}